\documentstyle[12pt,twoside,fleqn,espcrc1]{article}

\input psfig.sty
\def\lsim{\mathrel{\rlap{\lower4pt\hbox{\hskip1pt$\sim$}}
    \raise1pt\hbox{$<$}}}         
\def\gsim{\mathrel{\rlap{\lower4pt\hbox{\hskip1pt$\sim$}}
    \raise1pt\hbox{$>$}}}         

\newcommand{\AmS}{{\protect\the\textfont2
  A\kern-.1667em\lower.5ex\hbox{M}\kern-.125emS}}

\begin{document}
\title{The Canonical Nuclear Many-Body Problem as an Effective Theory}

\author{W. C. Haxton and T. Luu\address{
Institute for Nuclear Theory, Box 351550,
and Department of Physics, Box 351560, \\
University of Washington, Seattle, Washington  98195}}

\maketitle

\begin{abstract}
Recently it was argued that it might be possible to treat the conventional
nuclear structure problem -- nonrelativistic point nucleons interacting
through a static and rather singular potential -- as an effective
theory in a shell-model basis.  In the first half of this talk we
describe how such a program can be carried out for the simplest 
nuclei, the deuteron and $^3$He, exploiting a new numerical technique
for solving the self-consistent Bloch-Horowitz equation.  Some of 
the properties of proper effective theories are thus illustrated 
and contrasted with the shell model.  In the second half of the 
talk we use these examples to return to a problem that frustrated
the field three decades ago, the possibility of reducing the effective
interactions problem to perturbation theory.  
We show, by exploiting the Talmi integral expansion, that hard-core
potentials can be systematically softened by the introduction of
a series of contact operators familiar from effective field theory.
The coefficients of these operators can be run analytically by a renormalization group
method in a scheme-independent way, with the introduction of 
suitable counterterms.  Once these coefficients are run to the shell
model scale, we show that the renormalized coefficients contain all
of the information needed to evaluate perturbative insertions of
the remaining soft potential.  The resulting perturbative expansion
is shown to converge in lowest order for the simplest nucleus,
the deuteron.
\end{abstract}

\section{INTRODUCTION}
  
It is an honor to be able to address a distinguished audience at this celebration
for Achim Richter, a very good friend and valued colleague who has done so
so much to advance nuclear physics.  It is also great fun to return,
in this talk, to a topic that amused several of the theoreticians in
this room in the 1970's, the possibility of a perturbative expansion
of effective interactions and operators.  The thesis of this talk 
is that a marriage of the formidable technology of the shell model (SM)
with the modern ideas of the renormalization group
and effective theory (ET) may help us make some progress on this 
important problem.

In the first half of this talk we contrast the conventional
SM with the results of a proper ET carried
out in a SM basis.  We introduce a new and rather efficient numerical method
for solving the self-consistent Bloch-Horowitz equation governing
effective interactions and operators.  The method, while numerical,
systematically extracts from the excluded high-momentum space the
information needed to construct effective interactions,
and thus is in the spirit of ET techniques.  The results
for two simple nuclei, the deuteron and $^3$He, illustrate many
of the properties of correct effective theories (simple wave 
function evolution in the included space, nontrivial normalizations)
that are absent in the shell model.

The effective interactions problem is known to be highly nonperturbative.
In the second half of this talk we exploit these solutions in an
effort to understand the source of the nonperturbative scattering
at high momenta.  The first source
we encounter arises from the overbinding of harmonic oscillator
wave functions.  A straightforward reorganization of the Bloch-Horowitz
equation -- equivalent to resumming the kinetic energy operator to
all orders -- removes this problem.  The next source is hard-core
scattering.  To attack this problem we build a bridge between
effective field theories and potential problems by showing that
the contact operators introduced in the former generate a version of the 
Talmi integral expansion in a SM basis.  Thus a rather singular
NN potential can be systematically softened by removing the 
lowest order Talmi integrals, replacing these with contact 
potentials.  The effective interactions problem for these contact
interactions can be solved analytically via a shell-by-shell
renormalization group equation.  The hard-core effective interactions problem
corresponds to understanding the running of the coefficients of
these contact interactions.  The running can be done in a 
scheme-independent way -- that is, the effects of the hard core
on every matrix element in the SM space can be 
evaluated exactly -- with the introduction of a finite number
of higher-order contact interactions (or counterterms).  

With the hard core problem thus solved, we return to the remaining
soft parts of the potential, showing they can be inserted 
perturbatively among the infinite hard-core scattering series.
The only information needed is the renormalized coefficients
of the contact interactions.  We illustrate, for the lowest order
calculation in the deuteron, that this procedure reduces the
effective interactions problem to perturbation theory.

Although the talk focuses on the simplest possible example, the
techniques employed are quite general.  Details not presented here
can be found in \cite{Ha99B,Sav00,Luu00}.

\section{EFFECTIVE INTERACTIONS}
  
Consider a collection of nonrelativistic
nucleons interacting through a potential, e.g., of the Nijmegen or
Argonne/Urbana type, within an infinite Hilbert space
\begin{equation}
H = {1 \over 2} \sum_{i,j=1}^A (T_{ij} + V_{ij}),
\end{equation}
where $T_{ij}$ is the relative nonrelativistic kinetic energy
operator and $V_{ij}$ the nucleon-nucleon potential.  The related 
SM Hamiltonian acts in a restricted space and 
employs a softer ``effective" potential,
\begin{equation}
H_{SM} = {1 \over 2} \sum_{i,j=1}^A (T_{ij} + V^{eff}_{ij}).
\end{equation}
Motivating $H_{SM}$ is the notion that the determination of
$V^{eff}$ might be simpler than solving the original A-body problem:
the foundation of Brueckner theory is that high-momentum contributions
to the wave function might be integrated out in a rapidly converging
series in $\rho_{nuclear}$ or, equivalently, in the number of 
nucleons in high-momentum states interacting at one time outside
the SM space.

As will become apparent later, a typical SM space will 
contain explicitly $\sim$ 60\% of the
wave function that resides at long-wavelengths, thus guaranteeing at 
least a qualitative description of A-body correlations important to 
soft collective modes.  The strength of the SM is its capacity
to treat this long-wavelength part of the problem nonperturbatively:
the technology
for direct diagonalizations in large SM spaces is quite remarkable, including 
recent progress in Lanczos-based methods \cite{Ca98},
in treatments of light nuclei involving many shells \cite{Na98},
and in Monte Carlo sampling \cite{Ko97,Ho96}.  
Unfortunately the SM -- it is a model -- makes a number of 
uncontrolled approximations that distinguish it from an ET.
These are discussed in Ref.~\cite{Ha99B} and include effective
interactions lacking the correct functional form; a misunderstanding
of the proper normalization for SM wave functions; dependence of
results on fictitious parameters such as starting energies;
and the almost total neglect of effective operators.

\subsection{Self-consistent Bloch-Horowitz Solutions}
  
The effort to reformulate the SM as an ET begins with a definition
of the ``SM'' space.  The
goals of handling bound states and of generating an effective 
interaction that is translationally invariant leaves one sensible
choice:  many-body states constructed from harmonic oscillator
Slater determinants.  To exploit the relative/center-of-mass separability
of harmonic oscillator Slater determinants, one must separate the
SM and high-momentum spaces so that all configurations satisfying
\begin{eqnarray}
E \leq \Lambda_{SM} \hbar \omega_{SM}
\end{eqnarray}
are retained in the former.  For example, a SM calculation of 
$^{16}$O with
$\Lambda_{SM} = 4 + \Lambda_0$, where $\Lambda_0$ is the number of
quanta in the $^{16}$O closed shell, would include all
$4 \hbar \omega_{SM}$ configurations, e.g.,
$0p0h$, $2p2h$, and $4p4h$ excitations of nucleons from the $1p$ 
shell into the
$2s1d$ shell, $1p1h$ excitations of a $1s$ shell nucleon into 
the $3s2d1g$ shell, etc.  One can define the projection operator
onto the excluded high-momentum space by
\begin{equation}
Q_{SM} = Q(\Lambda_{SM},b_{SM}).
\end{equation}
where $b_{SM}$ is the oscillator parameter -- defined here in the
usual SM way for independent particle motion -- and
$\hbar \omega_{SM}={\hbar^2\over {M b_{SM}^2}}$is the corresponding energy. 
Thus the included or ``SM"
space is defined by two parameters, $\Lambda_{SM}$ and $b_{SM}$.  
The preservation of 
translational invariance is also important numerically, as it
reduces the two-body ladder to an effective one-body problem, etc.

The resulting Bloch-Horowitz (BH) equation \cite{Bl58} is then
\begin{eqnarray}
H^{eff} &=& H + H {1 \over E - Q_{SM}H} Q_{SM} H \nonumber \\
H^{eff} |\Psi_{SM} \rangle &=& E |\Psi_{SM} \rangle~~~|\Psi_{SM} \rangle
 = (1-Q_{SM}) |\Psi \rangle
\label{eq:BlHo}
\end{eqnarray}
where $|\Psi \rangle$ is the exact wave function and $H |\Psi \rangle
= E |\Psi \rangle$.  The difficulty posed by this equation is the
appearance of the unknown energy eigenvalue in the equation for
$H^{eff}$.  Thus this system must be solved self-consistently.
Note that there is no explicit reference to the harmonic oscillator
in this equation: it enters only implicitly through $Q_{SM}$ in
distinguishing the long-wavelength ``SM" space from the remainder
of the Hilbert space.
We emphasize that a proper solution of the BH equation must yield
results (energies and operator matrix elements) that are 
independent of $Q_{SM}$.

A new procedure for solving this problem was introduced in Ref.~\cite{Ha99B}.
It is based on a mapping of the full Hamiltonian in the high 
momentum space -- this space can be made finite, extending to
some scale $\Lambda_\infty \sim$ 3 GeV adequate to fully 
resolve the hard cores of realistic potentials --
into a simpler, truncated Hamiltonian via
the Lanczos algorithm, followed by the demonstration that the
the BH Green's function can be constructed from the Lanczos
matrix as a {\it function} of $E$, thus solving the effective
interactions problem.  That is, one can iterate on $E$ simply
by repeating a shell model calculation, as the effective interaction
can be immediately generated for any desired $E$.  In practice
the procedure converges very rapidly.

In our view the Lanczos approach to the effective interactions
problem appears to be remarkably simpler than the standard 
procedures of the field, particular in view of the need for
extensions to multi-nucleon ladders.  The traditional approach
divides the effective interactions problem into an energy-independent
piece (often called the Q-box) and an energy-dependent one
(represented by folded diagrams).  Frequently the energy 
dependence is removed by making a unitary transformation to
a non-Hermitian effective Hamiltonian.  The Lanczos procedure
is also much more in the spirit of EFT: the full Hamiltonian 
in the high-momentum space is never constructed.  Rather, it is
replaced by a much smaller Lanczos matrix which contains 
exactly the most relevant long-wavelength information of the
full matrix, the 2$n$-1 lowest moments (here $n$ represents the 
$n$th iteration in the Lanczos algorithm).  Thus the procedure 
can be viewed as a numerical ET in which this
information is recursively extracted.  Self-consistent solutions must be
obtained for each state, thus yielding a Hermitian but energy-dependent
effective interaction.
  
\subsection{Properties of Effective Wave Functions and Operators}
  
A first test of the techniques outlined above is to solve the
BH equation for some SM-like space to then see
if the resulting self-consistent energy is, indeed, the 
correct value (e.g., in agreement with Faddeev or other exact
calculations).  For model spaces of 2, 4, 6, and 8$\hbar \omega_{SM}$
in the case of the deuteron we obtained a binding energy of
-2.224 MeV (using $\sqrt{2}b_{SM}=1.6f$ and $\Lambda_\infty = 140$).  
The exact result is -2.2246 MeV.  Our ${}^3$He result for
same SM spaces and
$\Lambda_\infty = 60$ is -6.87 MeV, in agreement with the
corresponding Green's function Monte Carlo (GFMC) result of -6.87 $\pm$ 0.03 MeV.
Note that energies are variational in $\Lambda_\infty$, decreasing
as $\Lambda_\infty$ is increased.

More interesting is the evolution of the wave functions, 
shown in Table~\ref{table:two}, which is quite  
unlike that of typical shell model calculations.
The wave functions obtained in different model spaces agree 
over overlapping parts of their Hilbert spaces.  Thus as one
proceeds through 2$\hbar \omega_{SM}$, 4$\hbar \omega_{SM}$, 6$\hbar \omega_{SM}$,
... calculations, the ET wave function evolves only by adding 
new components in the expanded space.  The normalization of the
wave function grows accordingly.  (This normalization must be
calculated, as described below.)  Thus, for ${}^3$He,
the 0$\hbar \omega_{SM}$ ET calculation contains 0.311 of the full
wave function in the effective space; the 0+2+4$\hbar \omega_{SM}$
result is 0.700.  

Although there is clearly an intimate relation between effective
interactions and effective operators, it is standard
in the shell model to calculate nuclear responses with
bare operators, or perhaps with bare operators renormalized 
according to effective charges determined phenomenologically
at $q^2$ = 0, using SM wave functions normed to 1.
As we now have a series of exact effective interactions
corresponding to different model spaces, we can test
the validity of this approach.  The results for the elastic
magnetic form factors for the deuteron and $^3$He are shown in
Fig. 1.

\begin{table}
\caption{ET results for the ${}^3$He ground state wave function 
calculated with the Argonne $v18$ potential.
The columns on the
right correspond to different choices of the ET model space,
the analog of a SM space.  The rows correspond to the resulting amplitudes
for the designated, selected configurations.
The quantities
within the parentheses are the square of the norm of the 
effective wave function, e.g., the fraction of the ${}^3$He
ground state that resides in the ``SM'' space.
The basis states are
designated somewhat schematically as $\mid N, \alpha \rangle$,
where $N$ is the total number of oscillator quanta and 
$\alpha$ is an index representing all other quantum numbers.}
\label{table:two}
\vspace{0.2cm}
\begin{center}
\begin{tabular}{|r|r|r|r|r|r|r|}
\hline
\hline
 & \multicolumn{6}{c|}{amplitude} \\ \cline{2-7}
state & 0$\hbar \omega_{SM}$ & 2$\hbar \omega_{SM}$ &
4$\hbar \omega_{SM}$ & 6$\hbar \omega_{SM}$ & 8 $\hbar \omega_{SM}$ & exact \\ \cline{2-7}
 & (31.1\%) & (57.4\%) & (70.0\%) & (79.8\%) & (85.5\%) & (100\%) \\ \hline
 $\mid 0, 1 \rangle$ & 0.55791 & 0.55791 & 0.55791 & 0.55795 & 0.55791 & 
0.55793 \\ \hline
 $\mid 2, 1 \rangle$ & 0.00000 & 0.04631 & 0.04613 & 0.04618 & 0.04622 & 
0.04631 \\ \hline
 $\mid 2, 2 \rangle$ & 0.00000 & -0.48255 & -0.48237 & -0.48243 & -0.48243 & 
-0.48257 \\ \hline
 $\mid 2, 3 \rangle$ & 0.00000 & 0.00729 & 0.00731 & 0.00730 & 0.00729 & 
0.00729 \\ \hline
 $\mid 2, 4 \rangle$ & 0.00000 & 0.16707 & 0.16698 & 0.16706 & 0.16706 & 
0.16708 \\ \hline
 $\mid 4, 1 \rangle$ & 0.00000 & 0.00000 & -0.02040 & -0.02042 & -0.02043 & 
-0.02047 \\ \hline
 $\mid 4, 2 \rangle$ & 0.00000 & 0.00000 & 0.11267 & 0.11274 & 0.11275 & 
0.11289 \\ \hline
 $\mid 4, 3 \rangle$ & 0.00000 & 0.00000 & -0.04191 & -0.04199 & -0.04208 & 
-0.04228 \\ \hline
 $\mid 4, 4 \rangle$ & 0.00000 & 0.00000 & 0.28967 & 0.28978 & 0.28978 & 
0.29001 \\ \hline \hline
\end{tabular}
\end{center}
\end{table}

One sees in each case that by the time one 
reaches a momentum transfer $q \sim 2.5/f$, random numbers are
being generated: bare operators used in conjunction with exact
effective wave functions generate results that differ by an
order of magnitude, depending on the choice of the model space.
This is not surprising: if one probes the nucleus
at momentum transfers $\gsim 2k_F$, where $k_F$ is the Fermi 
momentum, most of the resulting amplitude should reside outside
the long-wavelength model space. 
That is, the strength resides entirely in the effective
contributions to the operator.  If these components are ignored,
the results have to be in error.

Clearly the effective interaction and effective operator have to
be treated consistently and on the same footing.  If $\hat{O}$
is the bare operator, one finds
\begin{equation}
\langle \Psi_f | \hat{O} | \Psi_i \rangle \equiv \langle \Psi_f^{eff}
 | \hat{O}^{eff} | \Psi_i^{eff} \rangle 
\end{equation}
where
\begin{equation} 
\hat{O}^{eff} = (1 + HQ_{SM} {1 \over E_f - HQ_{SM}}) \hat{O}
(1 + {1 \over E_i - Q_{SM}H}Q_{SM}H)
\end{equation}
and where the effective wave function normalization of 
$|\Psi_i^{eff} \rangle$ and $|\Psi_f^{eff} \rangle$, mentioned
earlier, must be determined using the effective operator $\hat{1}$,
e.g.,
\begin{equation}
1 = \langle \Psi_i | \Psi_i \rangle = \langle \Psi_i^{eff} |
(1 + HQ_{SM}{1 \over E_i - HQ_{SM}})(1 + {1 \over E_i - Q_{SM}H}
Q_{SM}H) | \Psi_i^{eff} \rangle
\end{equation}
These expressions can be evaluated with the Lanczos Green's function
methods described earlier.  When this is done, all of the effective calculations,
regardless of the choice of the model space, yield the same result,
given by the solid lines in Fig. 1.

We would argue, based on this example, that many persistent
problems in nuclear physics --- ranging from the renormalization
of $g_A$ in $\beta$ decay to the systematic differences
between measured and calculated M1 electromagnetic form factors ---
very likely are due to naive treatments of operators, treatments
that fail to satisfy the basic rules of ETs.  It
should be apparent from the above example that no amount of work
on $H^{eff}$ will help with this problem.  What is necessary is
a diagrammatic basis for generating $H^{eff}$ that can be 
applied in exactly the same way to evolving $\hat{O}^{eff}$.
{}From this perspective, phenomenological derivations of $H^{eff}$
by fitting binding energies and other static properties of 
nuclei are not terribly helpful, unless one intends to 
simultaneously find phenomenological renormalizations for 
each desired operator in each $q^2$ range of interest.

The point of this discussion has been largely pedagogical.
To our knowledge the examples given above are the only ones in
the classical nuclear physics literature in which a ``model
space" calculation has been formulated in a way that satisfies the 
basic rules of ET.  The results ---
particularly the invariance of energies and operator matrix
elements under changes in how we define $Q$ --- are obvious 
from the perspective of ET.  The fact that standard techniques
like the SM violate so many of the rules of ETs 
is cause for optimism: much can be done to improve the rigor of
such nuclear physics tools.  As we have demonstrated here, new 
numerical methods can be developed to handle the full ET problem 
at the cost of only modest additional effort.

\begin{figure}[!ht]
\hskip 1.0in\psfig{bbllx=0.cm,bblly=1.0cm,bburx=18cm,bbury=24.5cm,figure=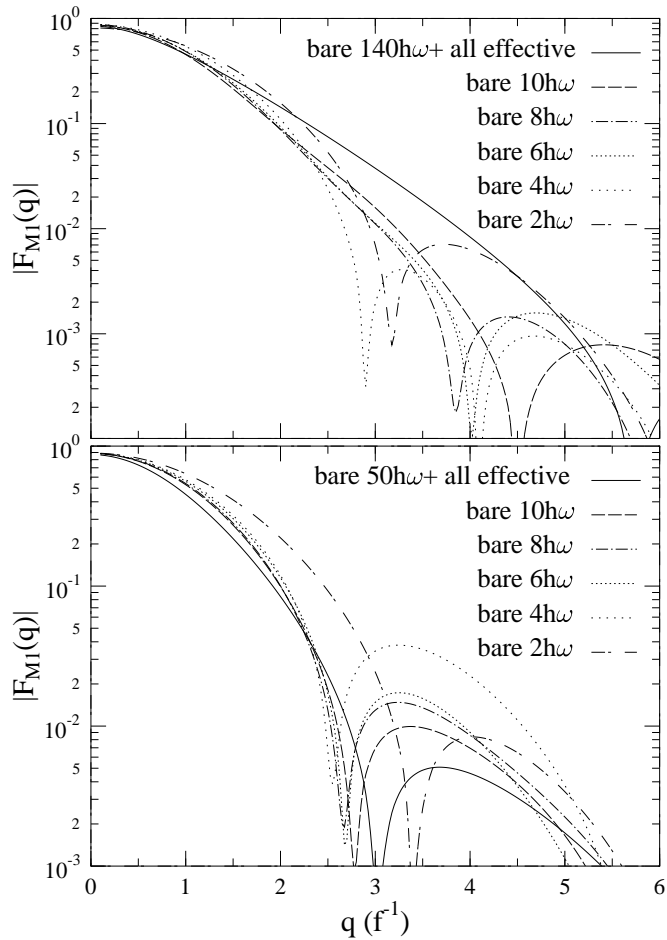,height=4.7in}
\caption{The magnetic elastic form factor for the deuteron (top)
and $^3$He (bottom)
calculated with the exact $H^{eff}$, SM wave functions normalized
to unity, and a bare operator are compared to the exact result
(solid line)~\protect\cite{Ha99A,Ha99B}.  
When effective operators and the proper wave
function normalizations are used, all results become identical
to the solid line.}
\end{figure}

\section{RESUMMATIONS: LONG AND SHORT DISTANCES}

The effort to derive SM effective interactions directly from 
realistic NN interactions was largely abandoned in the early 
1970's when it became apparent that the high-momentum scattering
problem was nonperturbative.  A series of efforts involving 
perturbative expansions in either the potential $V$ or the 
corresponding $G$-matrix (the sum of the two-body ladder diagrams)
yielded discouraging results, with higher-order terms often
dominating over lower-order ones\cite{Bar99}.  In particular
Shucan and Weidenmuller showed that any overlap in the spectra
of included and excluded, high-momentum states would lead to
a nonperturbative series\cite{Sh72,Sh73}.  Thus, the uncontrolled
approximations in the SM approach grew out of necessity: 
it became a largely phenomenological tool in part because 
early efforts to provide a more rigorous basis for the model
ended in failure.

Given our two SM-like ET calculations, the deuteron and ${}^3$He,
it is interesting to return to this old problem of the 
nonperturbative behavior of the nonrelativistic many-body
problem, using these examples as laboratories.   
Today's challenge, in fact, is more modest than that of the
1970's: modern computers coupled with new algorithms, like the
Lanczos solution of the BH equation introduced 
above, allow us to solve rather complex nonperturbative problems.
To extend the calculations above to ${}^4$He and heavier nuclei
one has to achieve only modest increases in efficiency,
such as a substantial lowering of $\Lambda_\infty$.  
Below we summarize some of our most recent work \cite{Luu00} that 
suggests that the goals of the 1970's --- a perturbative 
treatment of the nonrelativistic nuclear physics problem ---
may not be out of the question.  Through renormalization group 
and other techniques, the most severely nonperturbative contributions 
to the potential can be systematically extracted and treated.  In effect,
we try to show how many of the ideas of EFT can be adapted to the
nuclear potential problem, without abandoning the basic idea of
the SM that rigorous nonperturbative diagonalizations in a 
long-wavelength included space are important to the physics.

\subsection{Harmonic Oscillator Overbinding and Kinetic Energy Resummations}
  
We begin by studying the behavior of SM effective interaction matrix
elements $\langle \alpha | H^{eff} | \beta \rangle$  
in a perturbative expansion of the BH propagator
\begin{equation}
{1 \over E-QH} = {1 \over E-H_0} + {1 \over E-H_0} Q(V-V_0) {1 \over E-H_0} + \cdots,
\end{equation}
evaluating the series term by term.  Here $H_0$ and $V_0$ are the
harmonic oscillator Hamiltonian and potential.  
We thus lower $\Lambda_\infty$ and test whether states above 
$\Lambda_\infty$ can be treated perturbatively.

Once a scale of $\Lambda_\infty \sim 70$ is reached, striking 
differences appear in the rate of convergence of different matrix
elements.  All are greatly improved in first and second order
perturbation theory.  Those matrix elements well below the 
``boundary" at $\Lambda_\infty$ converge quickly to their 
correct values in this way.  However, those matrix elements
$\langle \alpha | H^{eff} | \beta \rangle$ where either
$| \alpha \rangle$ or $| \beta \rangle$ resides in the last
shell (with $N=\Lambda_\infty$) converge only very slowly
to the correct values, after the initial improvement in low-order
perturbation theory.  Typically $\sim 10^3$ orders of perturbation
theory are required to produce the correct value.
This clearly suggests that slow convergence is
associated with the relative kinetic energy operator $QT$ contribution
to $QH$, as the only transitions to states outside the Hilbert space
generated by this operator have $\Delta n$ = 1.
At large $r$ the strength of this transition becomes quite
large, $-Q V_0(r) | \alpha \rangle$, reflecting the
unphysical asymptotic behavior of harmonic oscillator wave
functions.  This amplitude 
propagates nonperturbatively as $(V-V_0)/H_0 \sim 1$.  Eventually
enough high-momentum harmonic oscillator wave functions are coupled
together to produce the softer asymptotic fall-off characteristic
of the correct bound-state wave function.

As discussed in Ref.~\cite{Luu00}, the required resummation is 
guided by the observation that the true potential, $V(r)$, falls off
properly at large $r$.  Thus a reorganization of the BH
equation in which the propagator is always sandwiched between
$V(r)$ should remove the unwanted propagation.  This leads to the
following recasting of the BH equation:
\begin{eqnarray}
\langle \alpha | H^{eff} | \beta \rangle &=& \langle \alpha | T | \beta \rangle +
(\langle \hat{\alpha} | - \langle \alpha |) E-T (| \hat{\beta} \rangle - | \beta \rangle) \nonumber \\
&& + \langle \hat{\alpha} | V + V {1 \over E-QH} QV | \hat{\beta} \rangle
\end{eqnarray}
where
\begin{equation}
| \hat{\alpha} \rangle = {E \over E-QT} | \alpha \rangle
\label{eq:kegf}
\end{equation}
If $\alpha$ and $\beta$ are not in the last shell, $| \hat{\alpha} \rangle = | \alpha \rangle$
and $| \hat{\beta} \rangle = | \beta \rangle$, so that the above 
rewriting of the BH equation reduces to the original form.
Otherwise, a modified wave function generated by Eq.~(\ref{eq:kegf})
must be used.  This rather simple resummation removes the worst --
though not the most interesting -- source of nonperturbative 
behavior in the BH equation.

It should be clear that a large reduction in the
scale $\Lambda_\infty$ at the cost of an evaluation of Eq.~(\ref{eq:kegf})
represents a tremendous savings: in place of a dense matrix $QH$
whose elements have to be evaluated numerically, we have a sparse
matrix $QT$ whose matrix elements are known analytically.  
For example, in the case of the deuteron, $QT$
is triadiagonal in the harmonic oscillator basis, allowing us to 
write the Green's function in a continued fraction expansion.

\subsection{Talmi Integrals, Renormalization Group, and the Hard Core}
  
Once the replacement $| \alpha \rangle \rightarrow | \hat{\alpha} \rangle$
is made, $\Lambda_\infty$ can be lowered to $\sim 40$ while maintaining
1 keV accuracy, if one works to third order in perturbation theory.
But with further lowering, errors arise in the perturbative
expansion that persist even if calculations are carried to very
high order.  At $\Lambda_\infty =$ 30, third order perturbation theory 
reduces $\sim$ 10\% errors in $\langle n'=1 l'=0 | H^{eff} | n=1 l=0 \rangle$
to $\sim$ 0.2\%.  But an error in excess of 0.1\% --- corresponding to
50 keV --- persists after 10
additional orders of perturbation.  Numerically one can verify that 
the nonperturbative tail is generated by the scattering at very
small $r$.  As expected, the nonperturbative contributions to $s-d$ matrix elements are 
much smaller than those for $\langle n'=1 l'=0 | H^{eff} | n=1 l=0 \rangle$
and other $s-s$ matrix elements.

We think this hard-core problem can be solved.  To keep the
presentation reasonable, below we concentrate primarily on the lowest
order (LO) treatment.  (The NLO and NNLO results are given in Ref.~\cite{Luu00}.)
One begins with the Talmi integral expansion for radial harmonic oscillator
matrix elements in the relative coordinate
\begin{eqnarray}
\langle n' l' | V_{sr}(r) | n l \rangle &=& \int^\infty_0
R_{n'l'}(r) V_{sr}(r) R_{nl}(r) r^2 dr \nonumber \\
&=& \sum_p B(n',l';n,l;p) I_p(b)
\label{eq:bcoef}
\end{eqnarray}
where
\begin{eqnarray}
I_p(b) &=& {2 \over \Gamma(p+3/2)} \int^\infty_0 e^{-r^2/b^2} V_{sr}(r) {r^{2p+2} dr \over b^{2p+3}},
\label{eq:talmi}\\ b &=& \sqrt{2} b_{SM}. \nonumber
\end{eqnarray}
The $B(n',l';n,l;p)$ are known analytically\cite{BM67}.
Here $V_{sr}(r)$ denotes the short-range contributions to the potential
(we return to this point below).  The integrals $I_p$ can be viewed
as a systematic expansion of the nonperturbative hard core in terms
of the parameter $(r_c/b)^2$, where $r_c$ is a distance associated
with the size of the hard core that remains unresolved at scale $\Lambda_\infty$.
Thus the LO term in the expansion is
\begin{equation}
I_{p=0}^{(0)} = {2 \over \Gamma(3/2)} {1 \over b^3} \int_0^\infty V_{sr}(r) r^2 dr
\label{eq:ip0}
\end{equation}
while in NLO one includes
\begin{eqnarray}
I_{p=0}^{(1)} &=& {2 \over \Gamma(3/2)} {1 \over b^3} \int_0^\infty V_{sr}(r) r^2 (1 - {r^2 \over b^2}) dr \nonumber \\
I_{p=1}^{(0)} &=& {2 \over \Gamma(5/2)}  {1 \over b^5} \int_0^\infty V_{sr}(r) r^4 dr .
\label{eq:ip1}
\end{eqnarray}
In NNLO one obtains
\begin{eqnarray}
I_{p=0}^{(2)} &=& {2 \over \Gamma(3/2)} {1 \over b^3} \int_0^\infty V_{sr}(r) r^2 (1 - {r^2 \over b^2} + {r^4 \over 2b^4}) dr \nonumber \\
I_{p=1}^{(1)} &=& {2 \over \Gamma(5/2)}  {1 \over b^5} \int_0^\infty V_{sr}(r) r^4 (1 - {r^2 \over b^2}) dr \nonumber \\
I_{p=2}^{(0)} &=& {2 \over \Gamma(7/2)}  {1 \over b^7} \int_0^\infty V_{sr}(r) r^6 dr .
\label{eq:ip2}
\end{eqnarray}
and so on.

Now the radial component of the $s$-wave potential $V_0 \delta({\bf r})$
is
\begin{equation}
V_\delta^{(0)}(r) = V_0 {1 \over 4 \pi r^2} \delta(r).
\end{equation}
Substituting this into Eq.~(\ref{eq:bcoef}) yields
\begin{equation}
\langle n' l'=0 | V_\delta^{(0)}(r) | n l=0 \rangle = B(n',0;n,0;0) I_0^\delta(b)
\end{equation}
where
\begin{equation}
I_0^\delta(b) = {2 \over \Gamma(3/2)} {1 \over 4 \pi b^3} V_0
\end{equation}
It follows that the contact interaction will produce the exact 
lowest-order Talmi integral contribution (Eq.~(\ref{eq:ip0})) to $V_{sr}$ provided
\begin{equation}
V_0 = 4 \pi \int_0^\infty V_{sr}(r) r^2 dr .
\end{equation}

Using a normalization that will be convenient later, we can then
rewrite $V_{sr}({\bf r})$ as
\begin{equation}
V_{sr}({\bf r}) = V_{sr}^{(1)}({\bf r}) + 
b^3 {\pi^2 \over 2} \hbar \omega a_0^{ss}(\Lambda_\infty) \delta({\bf r})
\end{equation}
where $\omega = {\omega_{SM} \over 2}$ and $V_{sr}^{(1)}$ is a new potential whose leading-order behavior
is of order $(r_c/b)^2$ relative to $V_{sr}$ (determined by subtracting
from the original Talmi integral expression for $V_{sr}$
the contribution from Eq.~(\ref{eq:ip0})).
The coefficient $a_0^{ss}(\Lambda_\infty)$ is a dimensionless 
coupling.
(Note that throughout this discussion, we have suppressed the 
spin and isospin of $V_{sr}({\bf r})$.  Thus $V_{sr}(r)$ represents
the radial function obtained after spin and isospin matrix elements
have been taken.  For example, the components of the $av18$ 
potential contributing to $s-s$ transitions is
\begin{equation}
V_{sr}({\bf r}) = V_1(r) + V_2(r) \vec{\tau}_1 \cdot \vec{\tau}_2 +
V_3(r) \vec{\sigma}_1 \cdot \vec{\sigma}_2 +
V_4(r) \vec{\sigma}_1 \cdot \vec{\sigma}_2 \vec{\tau}_1 \cdot \vec{\tau}_2
\end{equation}
so that
\begin{equation}
V_{sr}(r) = V_1(r)-3V_2(r)+V_3(r)-3V_4(r)
\end{equation}
for the deuteron ($l=0,s=1,t=0$).)

The above procedure is, in fact, general.  The NLO 
contribution (Eq.~(\ref{eq:ip1})) can be removed from $s-s$ wave deuteron matrix elements
of $V_{sr}({\bf r})$ by introducing a second, higher-order contact operator
\begin{equation}
a_2^{ss}(\Lambda_\infty) {1 \over 2} (\overleftarrow{\nabla}^2 \delta({\bf r}) +
\delta({\bf r}) \overrightarrow{\nabla}^2)
\end{equation}
where
\begin{equation}
a_2^{ss}(\Lambda_\infty) = {8 \over 3 \pi b^5 \hbar \omega}
\int_0^{\infty} V_{sr}(r) r^4 dr,
\end{equation}
while the NNLO contributions can be removed by
\begin{equation}
a_4^{ss}(\Lambda_\infty) (\overleftarrow{\nabla}^2 \delta({\bf r}) \overrightarrow{\nabla}^2 +
{3 \over 10} (\overleftarrow{\nabla}^4 \delta({\bf r}) + \delta({\bf r}) \overrightarrow{\nabla}^4))
\end{equation}
where
\begin{equation}
a_4^{ss}(\Lambda_\infty) = {2 \over 9 \pi b^7 \hbar \omega}
\int_0^{\infty} V_{sr}(r) r^6 dr.
\end{equation}
Likewise there are coeficients of additional contact operators
(denoted $a_2^{sd}$, $a_4^{sd}$, and $a_4^{dd}$ in Ref.~\cite{Luu00})
that remove the Talmi integral contributions to $s-d$ and $d-d$
matrix elements to NNLO. 

Up to this point we have simply rewritten the ``bare" potential ---
the potential that acts in the space defined by $\Lambda_\infty$ ---
in an entirely equivalent form, exploiting the Talmi integral expansion
and the fact that the unique translationally-invariant two-body
contact operators generate those Talmi integrals.  Now we consider,
in LO, the effects on $a_0^{ss}$ of integrating out the single 
shell at $\Lambda_\infty$, thereby mapping the original problem 
into an effective one with $\Lambda_\infty \rightarrow \Lambda_\infty - 2$.
The resulting BH equation instructs one to simply renormalize
the coefficient $a_0^{ss}(\Lambda)$ in LO \cite{Luu00},
\begin{equation}
a_0^{ss}(\Lambda-2) = a_0^{ss}(\Lambda) + {\Gamma({\Lambda+3 \over 2})
\over ({\Lambda \over 2})!} {a_0^{ss}(\Lambda)^2 \over
E_0 - (\Lambda + {3 \over 2}) - a_0^{ss}(\Lambda) {\Gamma({\Lambda+3
\over 2}) \over ({\Lambda \over 2})!}}
\label{eq:rgeq}
\end{equation}
where $E_0 = E/\hbar \omega$ is a dimensionless energy.  ($E$ is the
BH energy being determined selfconsistently.)
This leading-order RG equation is an {\it operator} equation,
properly correcting {\it every} matrix element in the new space
below $\Lambda_\infty - 2$.  Clearly this difference equation can
be used to run $a_0^{ss}(\Lambda_\infty)$ to any desired new scale,
including $\Lambda_{SM}$.

While the RG equation is a difference equation, the connections with
EFT become even clearer by introducing the natural definition of
the derivative
\begin{equation}
{d a_0^{ss}(\Lambda) \over d \Lambda} \equiv {a_0^{ss}(\Lambda) - a_0^{ss}(\Lambda-2) \over 2}.
\end{equation}
As will be apparent from results given below, the terms in the
denominator of Eq.~(\ref{eq:rgeq}) that compete with $\Lambda$ are
one power lower in $\Lambda$.  Thus for large $\Lambda$ the RG
equation can be written in a more conventional form
\begin{equation}
\Lambda {d a_0^{ss}(\Lambda) \over d \Lambda} \sim {1 \over 2}
\sqrt{{\Lambda \over 2}} a_0^{ss}(\Lambda)^2
\end{equation}
which has the solution
\begin{equation}
a_0^{ss}(\Lambda) \sim - \sqrt{{2 \over \Lambda}}.
\end{equation}
(The RG equation can also be integrated exactly, yielding a slightly
more complicated form.)
As $\Lambda$ is a dimensionless nonrelativistic energy, this is 
equivalent to running as $1/p$, where $p$ is the momentum, a 
result familiar from EFT.  Similarly, $a_2^{ss}$ runs as 
$\Lambda^{-3/2}$ and $a_4^{ss}$ as $\Lambda^{-5/2}$.

This procedure is followed through order $(r/b)^4$ in Ref.~\cite{Luu00},
which involves a series of coupled RG equations for $a_0^{ss}$,
$a_2^{ss}$, $a_4^{ss}$, $a_2^{sd}$, $a_4^{sd}$, and $a_4^{dd}$.
Beyond lowest order the running is no longer automatically scheme
independent, but independence can be restored by introducing a
finite number of higher derivative counterterms in NLO and NNLO.
We also stress that the analogy between EFT approaches, where 
the coefficients of contact terms are determined by fitting data,
and the Talmi integral expansion is very close.  In NNLO, for example, the
three lowest Talmi integrals completely determine the set of
matrix elements coupling the $n=1$ and $n=2$ oscillator shells.
We can fit our $ss$ {\it bare} contact potential to those matrix
elements, then use the RG equation to run this potential to the
SM scale, arriving at the same point as our EFT colleagues.  It is
thus apparent that the detailed short-range form of the bare potential
is irrelevant to shell-model-inspired effective theories:
the needed quantities are Talmi integrals over the short-range
part of the potential.  Any change in the detailed radial behavior
of the potential that preserves the value of the Talmi integrals
will also leave the low-energy theory unchanged.

\subsection{Perturbation Theory}
  
In the above discussion we were careful to carry through the subscript
$sr$ on our Talmi integrals, denoting the fact that, at the outset,
we divide the potential in some smooth way into a long-range piece
and into a short-range component.  It is clearly the short-range
component -- that rather singular part of the bare potential that is
unresolved at the scale $\Lambda_{SM}$ -- that should be represented
by the contact operator expansion.  And clearly, while we have just
solved the short-range effective interactions problem exactly, we 
are left with a gentler long-range potential
\begin{equation}
V_{lr}(\Lambda_\infty) = V - V_{sr}(\Lambda_\infty)
\end{equation}
whose effects have not been evaluated.  This potential, if we have
defined $sr$ properly, is a soft potential that can be handled very
well within the SM space: this is where we exploit the marvelous
technology of the SM.  But left over is a small contribution that
couples the SM and high momentum spaces.  How is this treated?

The procedure is an expansion of the BH propagator
\begin{equation}
QH = Q(H_0+V_\delta) + Q(V_{lr}-V_0)
\end{equation}
treating the second term perturbatively.  This corresponds to
insertions of the soft potential into an infinite series of hard-core
scatterings.  Thus the lowest order term is 
\begin{equation}
V_\delta^{eff} = V_\delta + V_\delta {1 \over E - (H_0 + Q V_\delta)}
Q V_\delta
\end{equation}
the solution of which is just our renormalized contact interaction.
The higher-order terms could be easily evaluated if we had a convenient
expression for the Green's function appearing above.  Such an
expression exists and depends quite simply on the
renormalized contact interaction coefficients.  The expressions
can be found in Ref.~\cite{Luu00}.

The net result for the deuteron ground state energy is shown in
Fig. 2.  This calculation employs a 10 $\hbar \omega$
SM space.  If one were to try to evaluate the SM effective
interaction perturbatively, the solid line is obtained.  The
answer oscillates rather wildly, order by order in perturbation theory,
failing to converge even after 20 orders of perturbation theory.
In contrast, the dashed line is the result of the procedure 
described here.  The oscillations damp quickly to an answer 
than is accurate to a few tens of kilovolts.  It appears, even 
in lowest order, that rather accurate results can be obtained
perturbatively.

\begin{figure}[!ht]
\hskip 0.5in\psfig{bbllx=0.0cm,bblly=1.5cm,bburx=17cm,bbury=14.5cm,figure=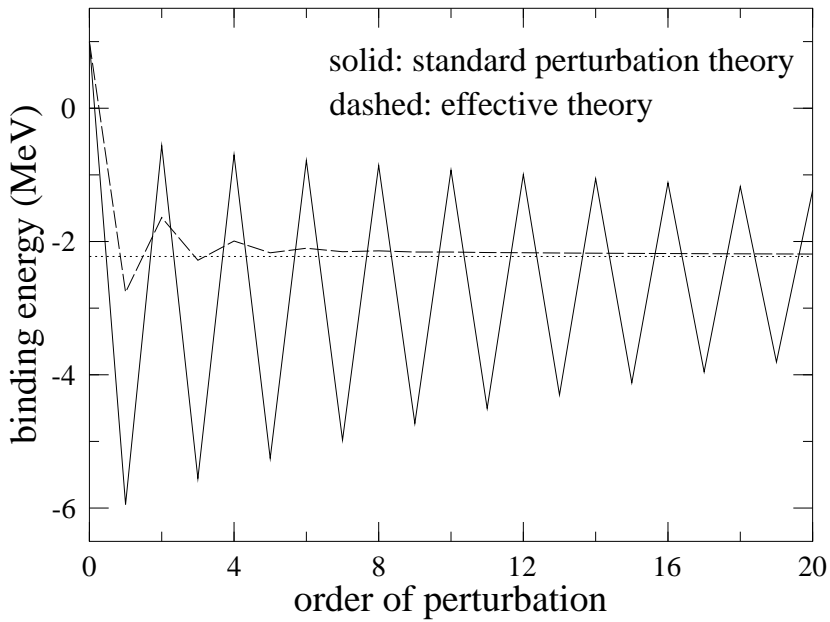,height=3.3in}
\caption{Calculations of the 
binding energy of deuterium in a $10 \hbar \omega_{SM}$
included space augmented by perturbation theory.
The solid result is a standard perturbation treatment around
the harmonic oscillator potential.  The 
answer is very slowly converging.  The dashed line is the 
ET result in which the hard core (defined in this application
as the potential within 0.7 f) is first treated via the LO renormalization
group procedure discussed in the text: the hard core is summed
to all orders analytically, and then the effects of the remaining
softer potential are evaluated perturbatively.  The convergence
improves remarkably.  The corresponding exact result ($\Lambda_\infty$ = 200)
is shown by the dotted line.}
\end{figure}

\section{OUTLOOK}
  
The extensions to three- and four-body clusters have not yet been
carried out, though nothing in the procedures is specific to the
deuteron.  The RG equations will, of course, involve  
the full set of general three- and four-body translationally 
invariant contact operators that can be constructed, just as 
EFT treatments of the three-body problem were required to
introduce three-body contact terms.  Thus the RG equations
become more complicated, but not conceptually more difficult.
If one began with a bare interaction that includes a short-ranged
three-body force, the coefficients of the three-body contact terms
would initially be nonzero, then evolve.  If there is no such force,
the initial value is zero, so that the three-body effective
interaction is then entirely induced.

If this program can be extended to heavier systems, some remarkable
simplifications might occur.  There exists no LO interaction that
involves more than four nucleons: at most four nucleons can exist
in a relative $s$-state.  Thus, the LO hard-core scattering problem
in Pb is no more difficult than the corresponding problem in 
${}^4$He, in some sense.  This is an explicit realization of 
Brueckner's idea of a cluster expansion at high momenta.
If such a system could be reduced to perturbation theory, as we
have achieved in the deuteron, some apparently unsolvable problems
(such as how to properly handle the projection operator $Q$ in
such systems) might be overcome.  Clearly there are many hurdles 
before one gets to this point, but we are cautiously optimistic that
more progress will be made.
  
In closing, it has been a great pleasure to speak at this
meeting is honor of a distinguished colleague and friend.  We
thank Achim Richter and the organizers for the opportunity to
discuss this work.
  
\section{ACKNOWLEDGEMENTS}
  
This work was supported in part by the US Department of Energy and
by the University of Washington through its Royalty Research
Foundation.  We thank our effective field theory colleagues,
particularly Martin Savage, for helpful discussions.

\end{document}